\documentclass[prl,aps]{revtex4}

\usepackage{graphicx}% Include figure files
\usepackage{dcolumn}% Align table columns on decimal point
\usepackage{bm}% bold math
\usepackage{color}
\usepackage{eso-pic}
\usepackage{amsmath}

\begin{document}

\title{Two-mode Correlation of Microwave Quantum Noise Generated by Parametric Down-Conversion}

\author{N. Bergeal$^{1,2}$, F. Schackert$^{2}$, L. Frunzio$^2$, M. H. Devoret$^2$}

\affiliation{$^1$Laboratoire de Physique et d'Etude des Mat\'eriaux - UMR8213-CNRS, ESPCI ParisTech, 10 Rue Vauquelin - 75005 Paris, France.}
\affiliation{$^2$Departments of Physics and Applied Physics, Yale University, New Haven, Connecticut, 06520-8284 USA.}
\date{\today}

  \begin{abstract}
  
 In this letter, we report the observation of the correlation between two modes of microwave radiation resulting from the amplification of quantum noise by the Josephson Parametric Converter. This process, seen from the pump, can be viewed as parametric down-conversion. The correlation is measured by an interference experiment displaying a contrast better than 99$\%$ with a number of photons per mode greater than two hundred and fifty thousand.  Dispersive measurements of mesoscopic systems and quantum encryption can benefit from this development.

 \end{abstract}
\maketitle
\indent In quantum physics, the no-cloning theorem \cite{wooters,scarani} forbids taking an unknown, but pure, quantum state and preparing an identical copy of that state while leaving the original intact. Allowing the original to be destroyed as the copy is produced leads to the process called quantum teleportation \cite{bennett}. An attempt  at multiple cloning of a superposition of states can only lead to a single superposition of clone products  of the original states.  Parametric down conversion is the simplest example of this by-pass of the no-cloning theorem \cite{yuen}. It is performed by letting the quantum noise of two signal channels enter the two input ports of a non-degenerate parametric amplifier. The two signals appearing at the output ports of the amplifier are strongly correlated for finite amplifier gain and become identical in the limit of infinite gain. From the point of view of the pump signal powering the amplifier, each of its photon is split by the non-linear component of the parametric amplifier into twin photons, which differ only in frequency and exit in separate channels. When measuring the photons by two ideal detectors, one would find rigorously the same number in both channels, even though, taken separately, each channel displays a Boltzmann statistics for the number of photons, as if it  was a spectral component of a thermal source. The peculiar entanglement of the two channels, often referred as two-mode squeezing \cite{yuen}, can be applied to quantum metrology measurements \cite{anisimov,glasser}, quantum cryptography protocols \cite{shapiro} and quantum teleportation \cite{milburn,furusawa}. So far, two-mode squeezing operation has been demonstrated exclusively in quantum optics experiments \cite{slusher}. In this letter, we report the observation of the correlation between two modes of microwave radiation resulting from the amplification of quantum noise by the Josephson parametric converter \cite{bergeal1,bergeal2}. Our work is motivated by the present interest in quantum information processing at microwave frequencies by superconducting integrated circuits \cite{manucharyan,lupascu,sillamaa,majer}. \\

Before presenting our experimental results, let us describe the properties of parametric down conversion in the quadrature representation, which is appropriate to ultra-low noise measurements of microwave signals in the quantum regime. Non-degenerate parametric amplifiers involve two distinct internal resonant modes A and B, whose frequencies $f_a$ and $f_b$ differs by at least the sum of the bandwidth of the two resonances. This is in contrast with degenerate parametric amplifiers, which operate with only one internal resonant mode and thus cannot produce two-mode squeezing. The non-degenerate amplifier is characterized by the input-output relations \cite{bergeal1}

 \begin{eqnarray}
\hat{a}_{\mathrm{out}}=[\cosh\lambda] \hat{a}_{\mathrm{in}}+[\sinh\lambda ]\hat{b}^\dagger_{\mathrm{in}}\\
\label{in1}
\hat{b}_{\mathrm{out}}=[\cosh\lambda] \hat{b}_{\mathrm{in}}+[\sinh\lambda] \hat{a}^\dagger_{\mathrm{in}}
\label{in2}
\end{eqnarray}

  \begin{figure}[t]
\centering
\includegraphics[width=12cm]{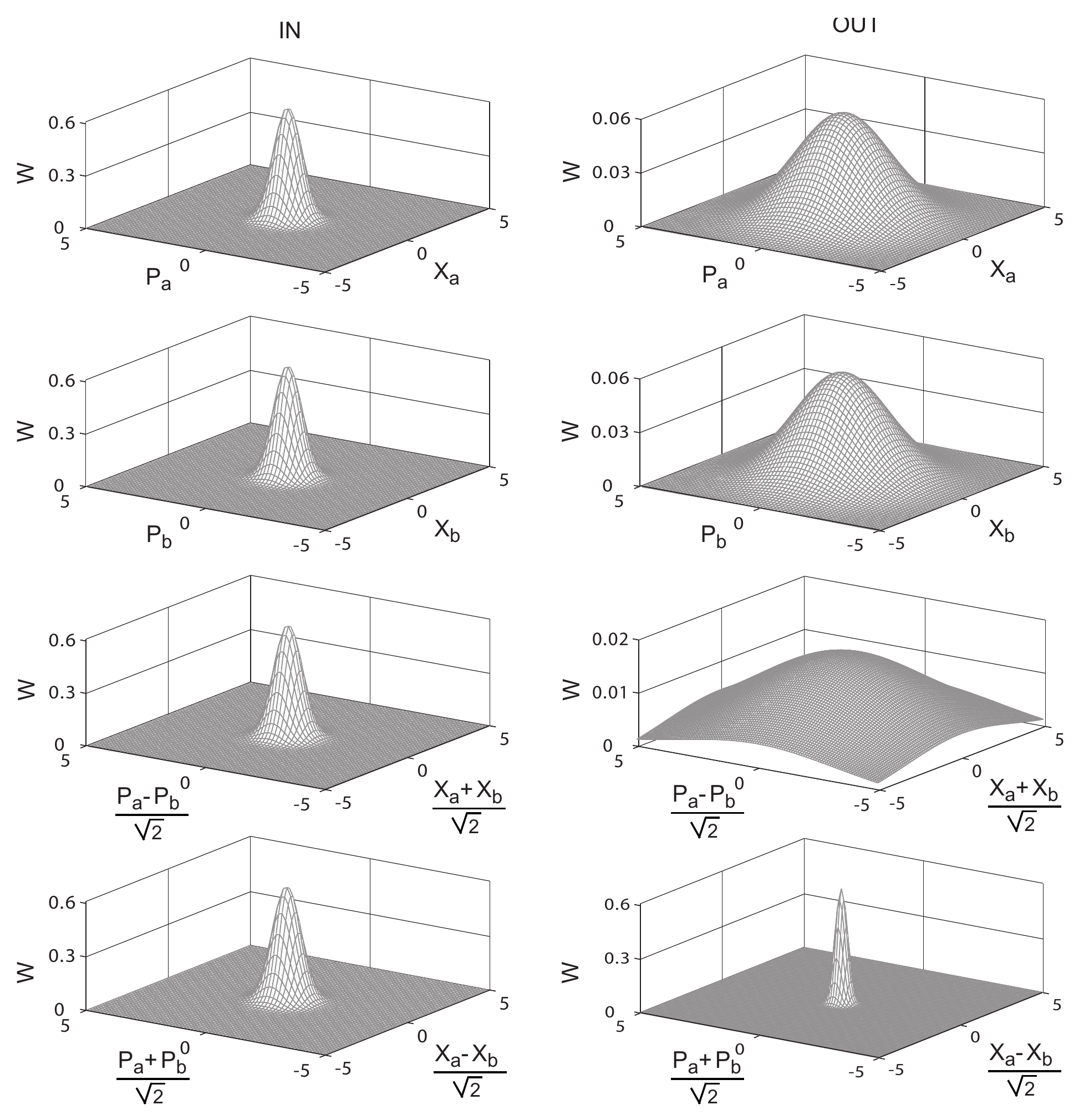}
\caption{Marginal distributions of the Wigner function at the input and output of a non- degenerate parametric amplifier represented in different quadratures. In this example $\lambda$=0.8}
\end{figure}  
where $\hat{a}$ ($\hat{a}^\dagger$) and $\hat{b}$ ($\hat{b}^\dagger$) denote the annihilation (creation) operators of the corresponding modes satisfying the commutation relation [$\hat{a}$,$\hat{a}^\dagger$]=[$\hat{b}$,$\hat{b}^\dagger$]=1, and where $\lambda$ is the squeezing parameter. The factors $\cosh(\lambda)$ and $\sinh(\lambda)$ play the role of the return and through gain for the amplifier, respectively. They increase monotonously with pump power until the threshold for parametric self-oscillation is reached. Squeezing and anti-squeezing correlations appear on the linear combinations 

 \begin{eqnarray}
\frac{\hat{a}_{\mathrm{out}}-\hat{b}^\dagger_{\mathrm{out}}}{\sqrt{2}}&=&\mathrm{e}^{-\lambda} \frac{\hat{a}_{\mathrm{in}}-\hat{b}^\dagger_{\mathrm{in}}}{\sqrt{2}}\\
\frac{\hat{a}_{\mathrm{out}}+\hat{b}^\dagger_{\mathrm{out}}}{\sqrt{2}}&=&\mathrm{e}^{\lambda} \frac{\hat{a}_{\mathrm{in}}+\hat{b}^\dagger_{\mathrm{in}}}{\sqrt{2}}
\end{eqnarray}

Consequently, when the input is in the vacuum state, at the output the quadrature component  $X_a^{\mathrm{out}}=(\hat{a}_{\mathrm{out}}+\hat{a}_{\mathrm{out}}^\dagger)/\sqrt{2}$ and   $X_b^{\mathrm{out}}=(\hat{b}_{\mathrm{out}}+\hat{b}_{\mathrm{out}}^\dagger)/\sqrt{2}$ are quantum correlated while $P_a^{\mathrm{out}}=-i(\hat{a}_{\mathrm{out}}-\hat{a}_{\mathrm{out}}^\dagger)/\sqrt{2}$  and  $P_b^{\mathrm{out}}=-i(\hat{b}_{\mathrm{out}}-\hat{b}_{\mathrm{out}}^\dagger)/\sqrt{2}$ are quantum anti-correlated. In other words, the variances  $(X_a^{\mathrm{out}}-X_b^{\mathrm{out}})^2=\mathrm{e}^{-2\lambda}(X_a^{\mathrm{in}}-X_b^{\mathrm{in}})^2$ and $(P_a^{\mathrm{out}}+P_b^{\mathrm{out}})^2=\mathrm{e}^{-2\lambda}(P_a^{\mathrm{in}}+P_b^{\mathrm{in}})^2$  are both smaller than the half photon of quantum noise. On the other hand, the variances  $(X_a^{\mathrm{out}}+X_b^{\mathrm{out}})^2=\mathrm{e}^{2\lambda}(X_a^{\mathrm{in}}+X_b^{\mathrm{in}})^2$ and  $(P_a^{\mathrm{out}}-P_b^{\mathrm{out}})^2=\mathrm{e}^{2\lambda}(P_a^{\mathrm{in}}-P_b^{\mathrm{in}})^2$ are larger than twice that of amplified quantum noise. 
The  output state  $|\Psi_\mathrm{out}\rangle$ is obtained by applying the unitary squeezing operator on the vacuum states of modes A and B \cite{barnett1,barnett2}

 \begin{eqnarray}
|\Psi^\mathrm{out}\rangle&=&\mathrm{e}^{\lambda\hat{a}^\dagger\hat{b}^\dagger-\lambda\hat{a}\hat{b}}|0\rangle_a|0\rangle_b\nonumber\\
&=&\frac{1}{\sqrt{1+\langle n \rangle}}\sum_{n=0}^\infty\bigg(\frac{\langle n \rangle}{1+\langle n \rangle}\bigg)^{\frac{n}{2}}|n\rangle_a|n\rangle_b
\label{photon}
\end{eqnarray}
 			 
where  $\langle n\rangle=\sinh^2\lambda$ denotes the mean photon number in each of the modes  at the output. This expression shows that the output state $|\Psi^\mathrm{out}\rangle$  is an entangled state in which the modes A and B contain the exact same number of photons. More directly linked to our experiment is the Wigner function associated with $|\Psi^\mathrm{out}\rangle$

  \begin{eqnarray}
 W(X_a^{\mathrm{in}},X_b^{\mathrm{in}},P_a^{\mathrm{in}},P_b^{\mathrm{in}})=\frac{1}{\pi^2}\exp\Big[\mathrm{e}^{2\lambda}\Big(\Big(\frac{X_a^{\mathrm{in}}+X_b^{\mathrm{in}}}{\sqrt{2}}\Big)^2+\Big(\frac{P_a^{\mathrm{in}}-P_b^{\mathrm{in}}}{\sqrt{2}}\Big)^2\Big)+\mathrm{e}^{-2\lambda}\Big(\Big(\frac{X_a^{\mathrm{in}}-X_b^{\mathrm{in}}}{\sqrt{2}}\Big)^2+\Big(\frac{P_a^{\mathrm{in}}+P_b^{\mathrm{in}}}{\sqrt{2}}\Big)^2\Big)\Big]
\end{eqnarray}
In figure 1, we compare the marginal distributions of the Wigner function at the input and output of the amplifier for different quadratures. As seen in fig 1a,b, each mode taken separately is transformed into the thermal state as shown by the reduced density operators 

\begin{eqnarray}
\rho_{a(b)} &=& \mathrm{Tr}_{{b(a)}}|\Psi^\mathrm{out}\rangle\langle\Psi^\mathrm{out}|\nonumber\\
&=&\frac{1}{1+\langle n\rangle}\sum_{n=0}^\infty\Big(\frac{\langle n\rangle}{1+\langle n\rangle}\Big)^n| n\rangle_{a(b)a(b)}\langle n|
\end{eqnarray}  
However, the squeezing and anti-squeezing are displayed in the joint quadrature components, as shown in fig 1c,d. The entanglement performed by the amplifier can be revealed by measuring the average intensity of an interference signal $\hat{c}$ obtained by superposing the output of one mode, say A, with the frequency shifted image of the output of the other mode, say B,
  \begin{eqnarray}
\hat{c}=\hat{a}_{\mathrm{out}}+\mathrm{e}^{i\phi}\hat{b}_{\mathrm{out}}^\dagger
\end{eqnarray}
				 
Here, $\phi$ is a phase shift applied to the pump signal at frequency $f_p = f_a + f_b$, used in mixing the frequency $f_b$ of the output of mode B, down to $f_a$. As $\phi$ is varied from 0 to $2\pi$, the average intensity $\langle\hat{c}^\dagger\hat{c}\rangle$ will exhibit a sinusoidal interference oscillation, whose minimal value is proportional to the variances of the squeezed quantities $X_a^{\mathrm{out}}-X_b^{\mathrm{out}}$ and $P_a^{\mathrm{out}}+P_b^{\mathrm{out}}$, and whose maximal value is proportional to the variances of the anti-squeezed quantities $X_a^{\mathrm{out}}+X_b^{\mathrm{out}}$ and $P_a^{\mathrm{out}}-P_b^{\mathrm{out}}$. This experiment thus accesses directly the variances of the marginal distribution of the Wigner function of fig 1.

  \begin{figure}[tb]
\centering
\includegraphics[width=12cm]{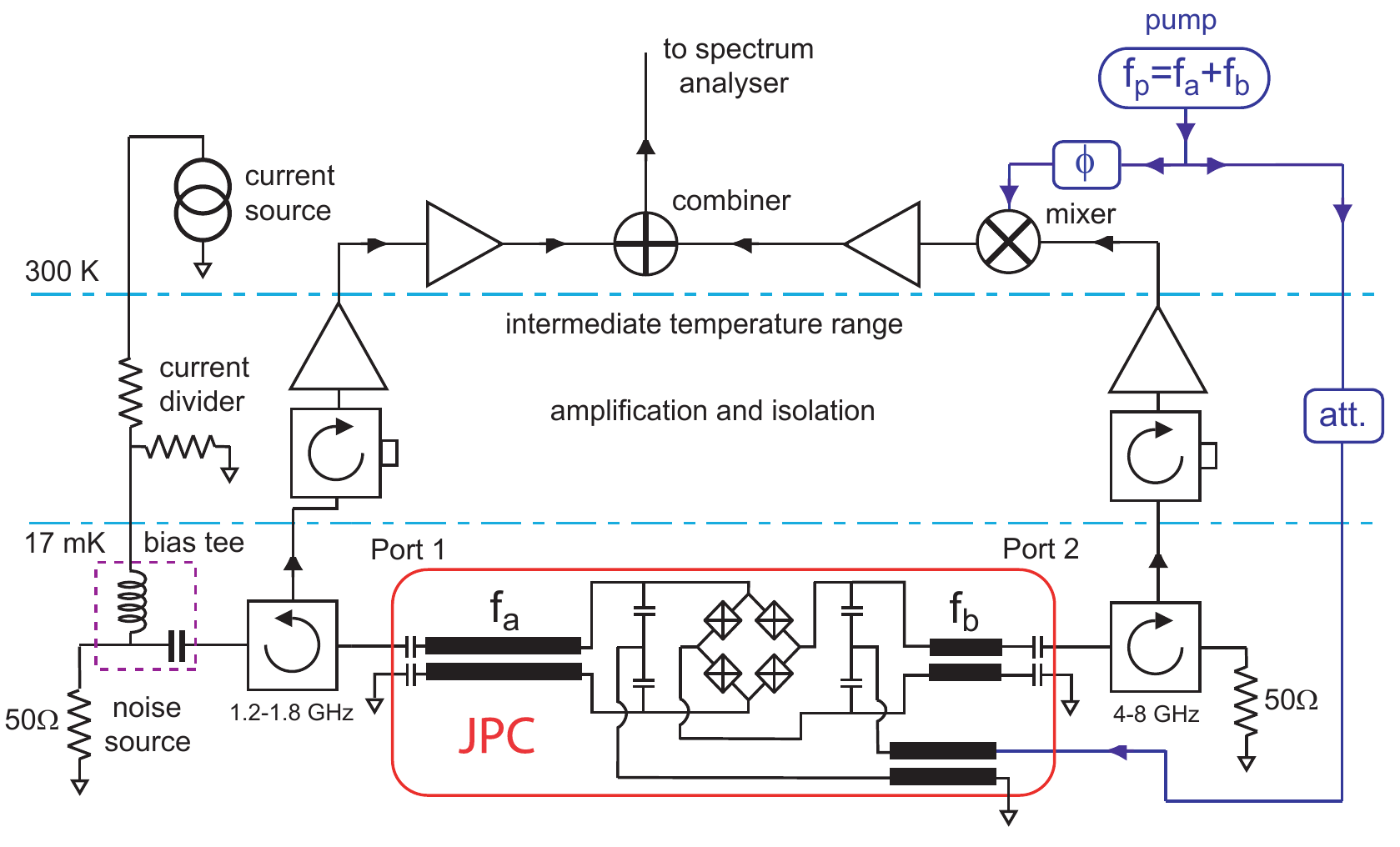}
\caption{The Josephson Parametric Converter (red solid line box) and its microwave measurement set-up. Port 1 is connected to a 50 $\Omega$ mesoscopic noise source whose effective temperature is controlled by a dc current source \cite{bergeal2}, and port 2 is connected to a 50 $\Omega$ load. The output signals are first amplified by HEMT cryogenic amplifiers at the 4.2K stage. A combination of isolators placed at different temperature stages are used to minimize the backaction of the amplifier on the JPC. At room temperature, the signal at $f_b$ is down converted to $f_a$ using a mixer, whose LO port is driven by the phase shifted ($\phi$) pump signal. Both signals are further amplified such that the total gains of the two amplification chains are identical, before being superposed by a combiner whose output is sent to a spectrum analyzer.}
\end{figure}

	The operation of our non-degenerate parametric amplifier, named Josephson Parametric Converter (JPC) (shown outlined in red in fig. 2) has been described in details in references \cite{bergeal1,bergeal2}. Its operation is based on the Josephson ring modulator, consisting of four nominally identical Josephson junctions forming a superconducting loop threaded by a magnetic flux $\phi$. Two superconducting resonators, defining the modes A and B, couple to the differential modes of the ring while being accessed by two external transmission lines.  An additional transmission line carries the pump signal at frequency $f_p$, and is weakly coupled to the common mode of the ring through a network of capacitors \cite{bergeal2}. In contrast with previous squeezing work involving degenerate Josephson parametric amplifiers, we have here a complete separation of the signal and idler modes both spatially and temporally \cite{yurke89, movshovich,lehnert,castellanos,yamamoto}. The JPC operates as a phase-preserving amplifier described by the characteristic input-output relations (1) and (2) where $\cosh\lambda=(1+\rho^2)/(1-\rho^2)$ and $\sinh\lambda=2\rho/(1-\rho^2)$, $\rho$ being the reduced pump current. A more general expression of return and through gain as function of input frequencies $f_1$ and $f_2$ can be found in reference \cite{bergeal2}.\\
	
	Our experimental set-up is described in figure 2. The two input ports of the JPC are connected to two 50 $\Omega$ loads anchored at the base temperature $T_0$=17 mK of a dilution refrigerator. Given the frequency $f_a$=1.6286 GHz and $f_b$=7.1694 GHz, and assuming thermal equilibrium in the loads, each port is in principle fed at its input with the half photon of quantum noise per mode. At the output of the JPC, the two output lines carry noise of complex amplitude $\hat{a}_{\mathrm{out}}$ and $\hat{b}_{\mathrm{out}}$ and frequencies $f_a$ and $f_b$.  These signals are amplified by a combination of cryogenic and room temperature amplifiers and become $g_a(\hat{a}_{\mathrm{out}}+\hat{a}_n)$ and $g_b(\hat{b}_{\mathrm{out}}+\hat{b}_n)$  where $\hat{a}_n$ and $\hat{b}_n$ are the complex amplitudes of the noise added by the amplifiers and $g_{a(b)}$ are the gain of the amplifier chains. Then the signal at $f_b$ is mixed with a signal at $f_p$, phase locked with the pump, producing a signal $g'_b(\mathrm{e}^{i\phi}\hat{b}^\dagger_{\mathrm{out}}+\hat{b}'_n)$, where $\phi$ is the phase shift between the LO port of the mixer and the pump. Note that the conjugation of $\hat{b}$ into $\hat{b}^\dagger$ in this operation, a crucial feature, is due to the fact that $f_p$ is greater than $f_b$, and that we are collecting the signal at $f_p-f_b$ and not the signal at $f_p+f_b$. The gains of the two chains have been adjusted to achieve the interferometer balance condition $g'_b= g_a$. The signals of the two channels are then superposed in a combiner whose output is sent to a spectrum analyzer, therefore performing the measurement of 
		 \begin{eqnarray}	
I(\lambda,\phi) =g_a^2\big\langle (\hat{a}_{\mathrm{out}}+\hat{a}_N+\mathrm{e}^{i\phi}\hat{b}_{\mathrm{out}}^\dagger+\hat{b}'_N) (\hat{a}_{\mathrm{out}}^\dagger+\hat{a}_N^\dagger+\mathrm{e}^{-i\phi}\hat{b}_{\mathrm{out}}+\hat{b}'^\dagger_N) \big\rangle
 \end{eqnarray} 

According to relations (1) and (2), the previous expression transforms into
  \begin{eqnarray}  
I(\lambda,\phi)&=&g^2_a\Big[\big\langle(X_a^\mathrm{in}+X_b^\mathrm{in})^2+(P_a^\mathrm{in}-P_b^\mathrm{in})^2\big\rangle\mathrm{e}^{2\lambda}\cos^2\frac{\phi}{2}
+\big\langle(X_a^\mathrm{in}-X_b^\mathrm{in})^2+(P_a^\mathrm{in}+X_b^\mathrm{in})^2\big\rangle\mathrm{e}^{-2\lambda}\sin^2\frac{\phi}{2}\Big]
+\langle \hat{a}_N\hat{a}^\dagger_N\rangle+\langle \hat{b}'^\dagger_N\hat{b}'_N\rangle\nonumber\\
&=&g_a^2\Big(\mathrm{e}^{2\lambda}\cos^2\frac{\phi}{2}+\mathrm{e}^{-2\lambda}\sin^2\frac{\phi}{2}+\langle \hat{a}_N\hat{a}^\dagger_N\rangle+\langle \hat{b}'^\dagger_N\hat{b}'_N\rangle\Big)
\label{inter}
 \end{eqnarray} 
 
Thus, $I(\lambda,\phi)$ displays two interference terms  $e^{2\lambda}\cos^2\Phi/2$ and $e^{-2\lambda}\sin^2\Phi/2$ , and in the limit of very large gain ($\lambda\gg1$) only the first survives, giving a full contrast to the oscillation of $I(\lambda,\phi)$ when varying $\phi$ from 0 to $\pi$.\\
	  
	  \begin{figure}[tb]
\centering
\includegraphics[width=14cm]{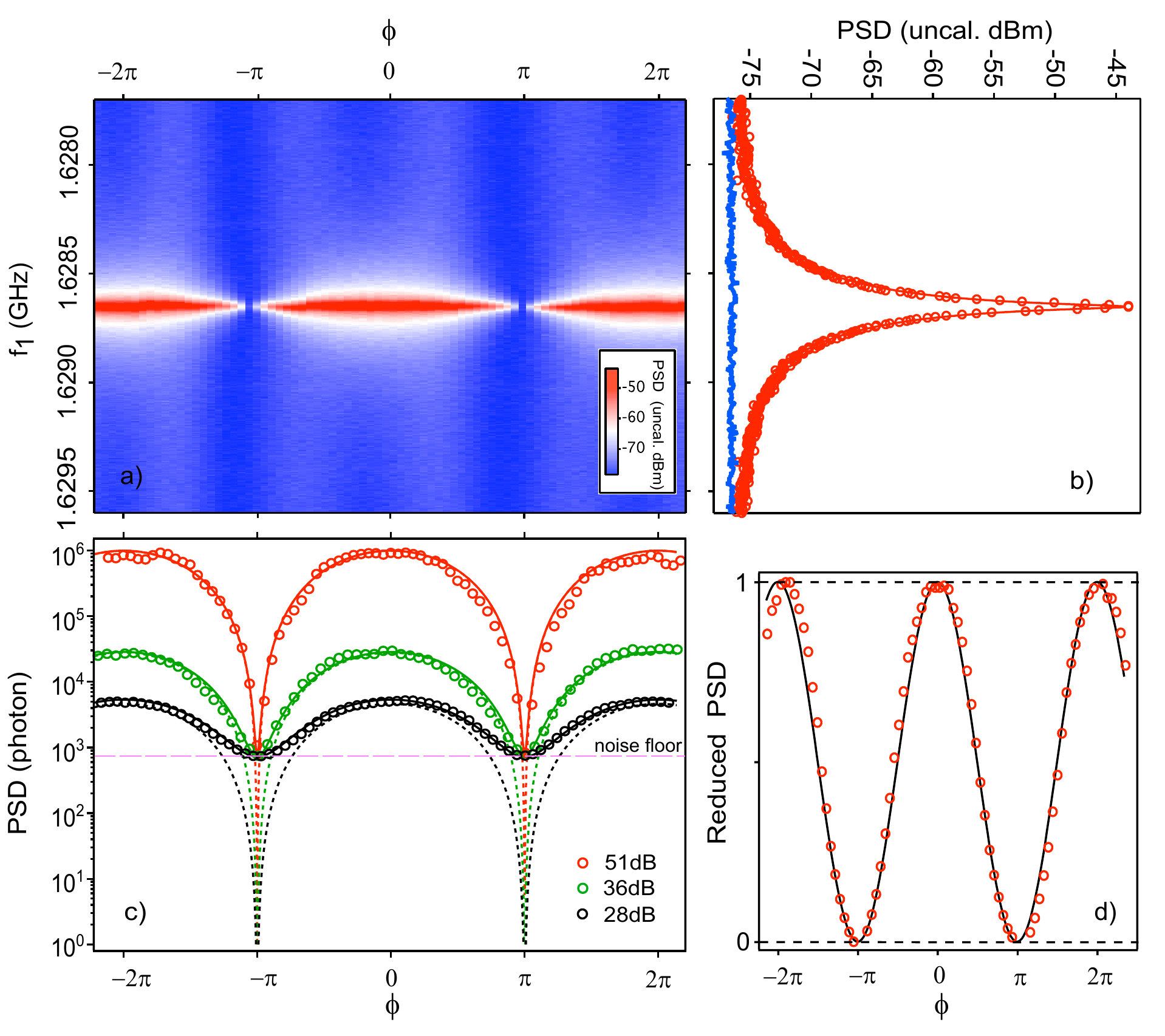}
\caption{a)  $I(\lambda,\phi)$ in color scale as a function of the noise frequency $f_1$ at port1 and the phase shift $\phi$ taken for a value of the gain of 51 dB. b) Cuts of $I(\lambda,\phi)$ corresponding to the constructive (maximum) and destructive (minimum) interference. c) $I(\lambda,\phi)$ integrated in its -3dB bandwith as a function of $\phi$, expressed in photon units referred to the output of the amplifier for three different values of gain. Dots corresponds to experimental data, full lines to the theoretical expression (10) and dashed lines to the theoretical expression (10) without the noised added by the chain amplifiers. d) Oscillations of the intensity of $I(\lambda,\phi)$  in a linear scale, integrated in  200 KHz band around $f_a$ showing a total contrast of the interference pattern. Dots corresponds to experimental data and the black line to a sinusoidal fit }
\end{figure}

	Figure 3 shows in color scale  $I(\lambda,\phi)$ measured as a function of the phase shift $\phi$ and the noise frequency $f_1$ for a power gain at the center of the band $G$=$\cosh^2\lambda$ of 51dB. The noise intensity is maximum at the center $f_a$ of the band and displays an interference pattern as function of $\phi$. As shown in figure 3b, the noise intensity drops down to the noise floor at $\phi$=(2n+1)$\pi$, which is set by the noise of the following amplifiers in the two signal chains and whose value is 32 dB lower than the peak intensity. In figure 3d, we plotted the sinusoidal oscillation of the intensity integrated in a 200 KHz band around the center frequency. Note that on the linear scale used for the figure, the background noise is invisible and the interference contrast is total. The minima of the interference for different gains are displayed in more details on figure 3c where the scale is given in photon units referred to the output of the amplifier, inferred from absolute noise measurements performed previously \cite{bergeal2}. They show that the destructive interference between the two noise channels can suppress down to $\approx$700 photons of the $\approx$ 500,000 photons (for the highest gain) corresponding to the sum of each channels taken separately (incoherent sum). However, given the JPC amplifier gain, the minima could reach in principle 2/$G$, but the present stability of the following amplifiers prevented us to measure this extreme squeezing effect governed by the second interference term in expression (\ref{inter}). The constructive interference corresponding to the anti-correlated state, produces twice as many photons as the sum of the each channel taken separately as expected.\\

The challenge of measuring the two-mode squeezing term in expression (\ref{inter}) could be addressed in an experiment with two JPC's with identical gain placed in series and where the interference would be produced by dephasing the two pump signals. Such experiment in which one of the arms would contain a mesoscopic system such as a qubit probed dispersively would constitute a quantum non-demolition measurement with no added noise. Alternatively, if a microwave photon detector would become available, one could directly measure the photon correlation expressed by relation (\ref{photon}).\\

We acknowledge useful discussions with R. Vijay, B. Huard,  N. Roch, S. M. Girvin and R. J. Schoelkopf. This
research was supported by the US National Security Agency through the US Army Research
Office grant W911NF-05-01-0365, the W. M. Keck Foundation, the US National Science Foundation
through grant No. DMR-0653377. L. F. acknowledges partial support from CNR-Istituto di Cibernetica. M. H. D. also acknowledges partial support from the College de France and from the French Agence Nationale
de la Recherche.

\end{document}